\documentclass{article}
\usepackage{spconf,amsmath,graphicx}
\usepackage{url}
\usepackage{multirow}
\usepackage{booktabs}
\usepackage[table]{xcolor}
\usepackage{subcaption}
\usepackage{hyperref}

\title{RGC-Bent: A Novel Dataset for Bent Radio Galaxy Classification}

\name{
\begin{tabular}{@{}c@{}}
Mir Sazzat Hossain$^{1,\star}$\thanks{$^\star$ Corresponding author: \href{mailto:mirsazzathossain@gmail.com}{mirsazzathossain@gmail.com}} \qquad 
Khan Muhammad Bin Asad$^{2,\dagger}$ \thanks{$^\dagger$ Expert Annotator.} \qquad 
Payaswini Saikia$^{3,\dagger}$\\
Adrita Khan$^{1,2}$\qquad
Md Akil Raihan Iftee$^1$\qquad 
Rakibul Hasan Rajib$^1$\qquad
Arshad Momen$^1$\\
Md Ashraful Amin$^1$\qquad
Amin Ahsan Ali$^1$\qquad
AKM Mahbubur Rahman$^1$
\end{tabular}}
\address{$^1$ Center for Computational \& Data Sciences, Independent University, Bangladesh\\
$^2$ Center for Astronomy, Space Science and Astrophysics, Independent University, Bangladesh\\
$^3$ Center for Astrophysics and Space Science, New York University Abu Dhabi}

\begin{document}

\maketitle 

\begin{abstract}
We introduce a novel machine learning dataset tailored for the classification of bent radio active galactic nuclei (AGN) in astronomical observations. Bent radio AGN, distinguished by their curved jet structures, provide critical insights into galaxy cluster dynamics, interactions within the intracluster medium, and the broader physics of AGN. Despite their astrophysical significance, the classification of bent radio AGN remains a challenge due to the scarcity of specialized datasets and benchmarks. To address this, we present a dataset, derived from a well-recognized radio astronomy survey, that is designed to support the classification of NAT (Narrow-Angle Tail) and WAT (Wide-Angle Tail) categories, along with detailed data processing steps. We further evaluate the performance of state-of-the-art deep learning models on the dataset, including Convolutional Neural Networks (CNNs) and transformer-based architectures. Our results demonstrate the effectiveness of advanced machine learning models in classifying bent radio AGN, with ConvNeXT achieving the highest F1-scores for both NAT and WAT sources. By sharing this dataset and benchmarks, we aim to facilitate the advancement of research in AGN classification, galaxy cluster environments and galaxy evolution. The  source code is available at: \url{https://github.com/mirsazzathossain/RGC-Bent}
\end{abstract}

\section{Introduction}
\label{sec:introduction}

Active Galactic Nuclei (AGNs) are some of the most luminous objects in the universe, powered by the accretion of matter onto supermassive black holes.
The classification of AGNs is important for understanding the evolution of galaxies in clusters, and the growth of supermassive black holes \cite{golden-marx2023}. Bent radio AGN (sometimes also called bent radio galaxies, head-tail galaxies, or bent-tailed galaxies) are a subset of radio-loud AGN, characterized by distinctly curved jet structures caused by interactions with the intergalactic medium \cite{rudnick1976}. These structures are commonly observed in galaxy clusters, where the interaction between the galaxy’s jets and the intracluster medium (ICM) plays a significant role in shaping their appearance. The bending of the jets is often linked to the relative motion between the host galaxy and the surrounding ICM, which results in ram pressure that distorts the jets’ trajectories \cite{Owen1976}.

\begin{figure}
    \centering
    \begin{subfigure}[b]{0.47\linewidth}
        \centering
        \includegraphics[width=\textwidth, keepaspectratio]{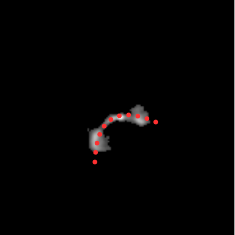}
        \caption{WAT}
        \label{fig:wat}
    \end{subfigure}
    \hfill
    \begin{subfigure}[b]{0.47\linewidth}
        \centering
        \includegraphics[width=\textwidth, keepaspectratio]{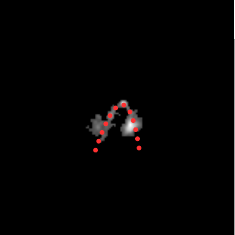}
        \caption{NAT}
        \label{fig:nat}
    \end{subfigure}
    \caption{Examples of Wide-Angle Tailed (WAT) and Narrow-Angle Tailed (NAT) radio galaxies. (a) WAT galaxies exhibit a wider ‘C’-shaped structure with jets extending over 90 degrees, while (b) NAT galaxies show a narrower ‘V’-shaped structure with jets within a 90-degree angle.}
    \label{fig:bent_agn}
    \vspace{-1.5em}
\end{figure}

Two notable subtypes are identified within this category based on the angle between their jets: Wide-Angle Tailed (WAT) Galaxies and Narrow-Angle Tailed (NAT) Galaxies, as shown in Figure \ref{fig:bent_agn}.
WAT galaxies feature two-sided jets that expand outward, creating a wide `C' shape with an opening angle exceeding 90 degrees~\cite{O'Dea2023}.
As the galaxy moves through the dense intracluster medium, interactions with it are believed to cause the bending of its jets.
In contrast, NAT galaxies, also called head-tail galaxies, have closely aligned jets forming a narrow `V' shape with an opening angle under 90 degrees and are usually located in galaxy clusters, moving at high velocities through the intracluster medium\cite{O'Dea2023, TernideGregory2017}.
The study of bent radio AGN, including WATs and NATs, is crucial for understanding the dynamics of galaxy clusters, the behavior of the intracluster medium, the interactions between galaxies and their environments, and the dynamics of AGN jets and their surrounding conditions \cite{golden-marx2023,Lao2025}.

In this work, we make the following contributions:
\begin{itemize}
    \vspace{-0.5em}
    \item We present a dataset of bent radio AGN images, captured by the Very Large Array (VLA), specifically compiled to facilitate machine learning-based classification.
    \vspace{-1.5em}
    \item We evaluate the performance of five machine learning models on this dataset, establishing a baseline for automated bent radio AGN classification.
    \vspace{-0.5em}
    \item We release all code as open source for key data preprocessing tasks, including background estimation, source identification, mask generation, and background removal via masking.
\end{itemize}

\vspace{-0.5em}
\section{Existing Datasets and Limitations}
\label{sec:existing_datasets}

Machine learning has been widely applied in astronomy, but the lack of specialized datasets remains a challenge. Existing datasets like Galaxy Zoo~\cite{lintott2008} focus on general galaxy classification and radio galaxy Zoo deals with radio sources~\cite{Wong2025}.
The MiraBest dataset~\cite{porter2023} provides labeled Fanaroff-Riley Type I and II sources~\cite{fanaroff74} but lacks the labeling of bent AGNs into WATs and NATs.

While no dedicated machine learning dataset exists for bent radio AGNs, several catalogs provide relevant information.
The Proctor et al.~\cite{Proctor2011} catalog was among the first, but it contained automated annotations which cannot be relied upon for high-precision tasks without manual verification.
In contrast, Sasmal et al.~\cite{sasmal2022} introduced a manually curated catalog, offering more accurate source identification.
This catalog serves as the primary data source for our dataset creation.
Converting it into a machine learning-ready dataset requires careful data acquisition, preprocessing, and expert validation.
In this paper, we document this process in detail to facilitate future research in bent radio AGN classification.
\vspace{-0.3em}

\section{Dataset Creation}
\label{sec:dataset}

\subsection{Data Acquisition}
The raw data is obtained from the Faint Images of the Radio Sky at Twenty-cm (FIRST) survey using source coordinates from the catalog proposed by Sasmal et.al.~\cite{sasmal2022}. The FIRST survey, conducted at 1.4 GHz with the Karl G. Jansky Very Large Array (VLA) in New Mexico, USA, covers the sky north of $\delta = -40^{\circ}$ with a 5-arcsecond resolution and a 0.15 mJy sensitivity.

The data is stored in Flexible Image Transport System (FITS) format and includes flux density measurements at different frequencies. We retrieve the images from NASA SkyView\footnote{\url{https://skyview.gsfc.nasa.gov}}, a virtual observatory providing access to multi-survey astronomical data. Using the provided source coordinates, we query SkyView to download the corresponding FIRST survey radio images, which are then stored in FITS format for subsequent processing.

\subsection{Data Processing}

The first few steps (up to and including Section \ref{pybdsf}) of data processing were performed using the Python package \texttt{PyBDSF} \cite{mohan2015}.

\subsubsection{Background Estimation}
The first step in processing is estimating the background noise level for each image. We compute the background level and its standard deviation by segmenting the image into smaller subregions and calculating the local mean (\(\mu_{\text{local}}\)) and standard deviation (\(\sigma_{\text{local}}\)) of the pixel intensities within each subregion. The background estimation is modeled as:
\vspace{-0.4em}
\begin{equation}
    B(x, y) = \mu_{\text{local}} + k \cdot \sigma_{\text{local}},
    \vspace{-0.4em}
\end{equation}
where \(B(x, y)\) denotes the local background intensity, and \(k\) is a scaling factor that accounts for typical noise fluctuations.

\subsubsection{Source Identification}
After background estimation, sources are identified by detecting contiguous regions (or “islands”) where the intensity significantly exceeds the local background. Two thresholds are applied during this process:
\begin{itemize}
    \item \textbf{Island Threshold (\(T_{\text{isl}}\)):} This parameter, typically set to 3, defines the minimum signal level for considering a group of contiguous pixels as part of an island. A pixel is included in an island if:
    \vspace{-0.4em}
    \[
    I(x, y) > B(x, y) + T_{\text{isl}} \cdot \sigma_{\text{local}},
    \vspace{-0.4em}
    \]
    helping to delineate regions where emission is likely associated with a source.
    \item \textbf{Peak Threshold (\(T_{\text{pix}}\)):} While \(T_{\text{isl}}\) defines the overall extent of an island, \(T_{\text{pix}}\), typically set to 5, establishes the detection threshold for source peaks. Only islands with peaks exceeding:
    \vspace{-0.4em}
    \[
    I(x, y) > B(x, y) + T_{\text{pix}} \cdot \sigma_{\text{local}},
    \vspace{-0.4em}
    \]
    are considered valid detections.
\end{itemize}

After thresholding, contiguous pixels satisfying these criteria are grouped into islands. For each island, a two-dimensional Gaussian model is fitted to extract key source parameters, including the centroid \((x_0, y_0)\), peak intensity \(I_0\), and morphological descriptors (major axis \(\sigma_{\text{maj}}\), minor axis \(\sigma_{\text{min}}\), and position angle \(\theta\)). The Gaussian model is given by:
\begin{equation}
    I(x, y) = I_0 \exp\left(-\frac{(x' - x_0)^2}{2 \sigma_{\text{maj}}^2} - \frac{(y' - y_0)^2}{2 \sigma_{\text{min}}^2}\right),
\end{equation}
where \((x', y')\) are the coordinates rotated by the angle \(\theta\). This model effectively characterizes the morphology of each source and helps distinguish genuine sources from noise artifacts.

\begin{figure}
    \centering
    \begin{subfigure}[b]{0.325\linewidth}
        \centering
        \includegraphics[width=\textwidth, keepaspectratio]{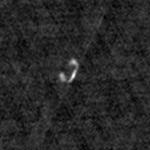}
        \caption{Original Image}
        \label{fig:original_image}
    \end{subfigure}
    \hfill
    \begin{subfigure}[b]{0.325\linewidth}
        \centering
        \includegraphics[width=\textwidth, keepaspectratio]{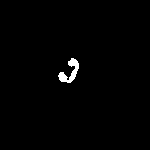}
        \caption{Generated Mask}
        \label{fig:generated_mask}
    \end{subfigure}
    \hfill
    \begin{subfigure}[b]{0.325\linewidth}
        \centering
        \includegraphics[width=\textwidth, keepaspectratio]{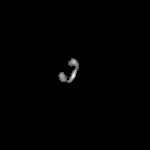}
        \caption{Masked Image}
        \label{fig:masked_image}
    \end{subfigure}
    
    \caption{Masking process overview. (a) Original image, (b) generated mask, and (c) resulting (final) masked image.}
    \label{fig:source_masking}
    \vspace{-1.0em}
\end{figure}

\begin{table}[!ht]
    \centering
    \caption{Distribution of NAT and WAT sources across training and test batches. The first nine batches (576 images) form the training set, while the last batch (63 images) serves as the test set.}
    \vspace{5pt}
    \label{tab:source_params}
    \setlength{\tabcolsep}{0.815em}
    \begin{tabular}{cccccc}
        \toprule
        \multirow{2}{*}{Batch} & \multicolumn{2}{c}{Source Count} & \multirow{2}{*}{Total} & \multirow{2}{*}{Cumulative Total} \\
        \cmidrule(lr){2-3}
        No. & WAT & NAT & Sources & (Train/Test) \\
        \midrule\noalign{\vskip -2.5pt}
        \rowcolor[gray]{0.9}\multicolumn{5}{c}{Training Set} \\
        \noalign{\vskip -1.5pt}\midrule
        0 & 39 & 25 & 64 & \multirow{9}{*}{576} \\
        1 & 39 & 25 & 64 & \\
        2 & 39 & 25 & 64 & \\
        3 & 39 & 25 & 64 & \\
        4 & 39 & 25 & 64 & \\
        5 & 38 & 26 & 64 & \\
        6 & 38 & 26 & 64 & \\
        7 & 38 & 26 & 64 & \\
        8 & 38 & 26 & 64 & \\
        \midrule\noalign{\vskip -2.5pt}
        \rowcolor[gray]{0.9}\multicolumn{5}{c}{Test Set} \\
        \noalign{\vskip -1.5pt}\midrule
        9 & 38 & 25 & 63 & 63 \\
        \midrule\noalign{\vskip -2.5pt}
        \rowcolor[gray]{0.9}\textbf{Total} & 385 & 254 & 639 & 639 \\
        \noalign{\vskip -1.5pt}\bottomrule
    \end{tabular}
    \vspace{-1.5em}
\end{table}

\begin{table*}[!ht]
    \caption{Quantitative evaluation (performance comparison) of different deep learning models on the bent radio AGN dataset. The table presents the accuracy of the models, the class-wise precision, recall, and F1-score for NAT and WAT sources. ConvNeXT achieves the highest F1-scores for both WAT and NAT sources, indicating its effectiveness in identifying both morphologies.}
    \label{tab:quantitative}
    \centering
    \setlength{\tabcolsep}{1.1em}
    \begin{tabular}{lccccccc}
        \toprule
        \multirow{2}{*}{Model} & \multirow{2}{*}{Accuracy [\%]} &\multicolumn{3}{c}{\textbf{WAT}} & \multicolumn{3}{c}{\textbf{NAT}} \\
        \cmidrule(lr){3-5} \cmidrule(lr){6-8}
        & & \textbf{Precision} & \textbf{Recall} & \textbf{F1-score} & \textbf{Precision} & \textbf{Recall} & \textbf{F1-score} \\ 
        \midrule \midrule
        VGG-16~\cite{simonyan2015}       & 74.60 & 0.73 & 0.92 & 0.81 & 0.80 & 0.48 & 0.60\\
        ResNet-50~\cite{he2016}          & 77.77 & 0.79 & 0.87 & 0.82 & 0.76 & 0.64 & 0.70\\
        ViT-B-16~\cite{dosovitskiy2021}  & 76.19 & 0.76 & 0.89 & 0.82 & 0.77 & 0.73 & 0.74\\
        SWIN-B~\cite{liu2021}            & 80.95 & 0.80 & 0.92 & 0.85 & 0.84 & 0.64 & 0.73\\
        ConvNeXT~\cite{liu2022}          & 84.12 & 0.85 & 0.89 & 0.87 & 0.83 & 0.76 & 0.79\\
        \bottomrule
    \end{tabular}
    \vspace{-1.0em}
\end{table*}

\subsubsection{Mask Generation}\label{pybdsf}
Once sources are detected and characterized, a binary mask is generated to delineate the spatial extent of each source. The mask \(M(x,y)\) is produced by thresholding the fitted Gaussian model:
\begin{equation}
    M(x, y) = 
    \begin{cases}
        1, & \text{if } I(x, y) > B(x, y) + T_{\text{pix}} \cdot \sigma_{\text{local}}, \\
        0, & \text{otherwise.}
    \end{cases}
\end{equation}
This criterion ensures that only regions with intensities significantly above the background are included in the mask. For sources exhibiting extended or diffuse emission, a dilation operation is applied to the mask. Dilation expands the mask boundary by some pixels \(d\), ensuring complete coverage of the source.

For most sources, the default parameters in \texttt{PyBDSF} are \(T_{\text{isl}} = 3\), \(T_{\text{pix}} = 5\), and a dilation parameter \(d = 0\) pixels. As a supplementary material, we give the list of sources for which these parameters were modified, along with the corresponding values of \(T_{\text{isl}}\), \(T_{\text{pix}}\), and \(d\). The final binary masks are stored in FITS format for subsequent analysis. The next steps were not implemented using PyBDSF.

\subsubsection{FITS to PNG Conversion}
To facilitate visualization and further processing, both the FITS images and the corresponding binary masks are converted to Portable Network Graphics (PNG) format. The pixel values in the FITS images are linearly scaled to the range [0, 255] using the following transformation:
\begin{equation}
    \text{img}_{\text{PNG}} = \frac{\text{img} - \min(\text{img})}{\max(\text{img}) - \min(\text{img})} \times 255,
\end{equation}
where \(\text{img}\) denotes the original FITS image, and \(\min(\text{img})\) and \(\max(\text{img})\) are the minimum and maximum pixel values in the image, respectively.

\subsubsection{Background Removal via Masking}
Background removal from the radio images is achieved by applying the binary masks generated in the previous step. By multiplying the original image with its corresponding binary mask, the background is suppressed, retaining only the source emissions. This operation is expressed as:
\begin{equation}
    \text{img}_{\text{masked}} = \text{img} \times M,
\end{equation}
where \(\text{img}_{\text{masked}}\) is the background-subtracted image, \(\text{img}\) is the original radio image, and \(M\) denotes the binary mask.

This process is applied to all images in the dataset, resulting in a series of background-subtracted images that highlight the radio sources. Figure \ref{fig:source_masking} illustrates the background removal process for a sample radio image.

\subsection{Final Dataset Curation}

\subsubsection{Expert Validation}
After processing the data, two expert astronomers from our group manually reviewed each image in the dataset to verify the presence of bent AGNs, the validity of the binary masks, the quality of the background removal, and the overall image quality. They annotated each image with a binary label indicating whether the source was suitable for inclusion in the dataset. In cases of poor-quality samples, the experts provided feedback explaining the reasons for exclusion. Only images unanimously approved by both experts were included in the final dataset. Consequently, 64 sources were rejected, leaving a total of 639 bent radio AGNs in the dataset.

\vspace{-0.5em}

\subsubsection{Dataset Splitting}
The dataset was split into 10 batches, each containing 64 images, except for the last batch, which contained 63 images. The dataset was stratified based on the source type (NAT or WAT) to ensure an even distribution of source types across the batches. Each batch was saved as a separate pickle file for easy loading during training and testing. The first nine batches were used for training, and the last batch was reserved for testing. Table \ref{tab:source_params} summarizes the distribution of sources across the batches.

\section{Benchmarking Experiments}
\label{sec:experiments}

\begin{figure*}[ht]
    \centering
    \includegraphics[width=\linewidth]{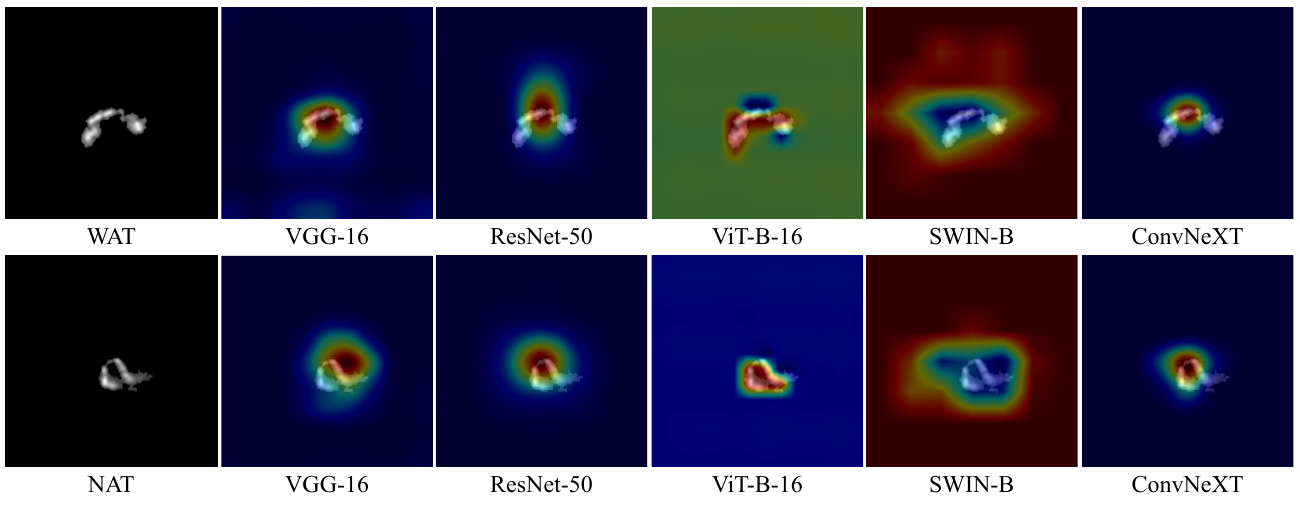}
    \caption{Class activation maps (CAMs) generated by different models for a sample NAT and WAT source using Grad-CAM~\cite{selvaraju2017}. The CAMs highlight the regions in the image that contribute most to the model's classification decision.}
    \label{fig:grad_cam}
    \vspace{-1.2em}
\end{figure*}

\subsection{Experimental Setup}

\subsubsection{Baseline Models}
We evaluate the performance of several state-of-the-art deep learning models for image classification. These include two widely used convolutional neural networks (CNNs), VGG-16~\cite{simonyan2015} and ResNet-50~\cite{he2016}, along with two transformer-based architectures, Vision Transformer (ViT)~\cite{dosovitskiy2021} and SWIN Transformer~\cite{liu2021}. Additionally, we consider ConvNeXT~\cite{liu2022}, a recent CNN-based model that has demonstrated competitive results on various image classification tasks. All models are pre-trained on ImageNet and fine-tuned on our dataset.

\vspace{-0.3em}

\subsubsection{Implementation Details}
The models are implemented using the PyTorch deep learning framework and trained on the curated dataset. The dataset consists of grayscale images of size $150 \times 150$ pixels, which are resized to $224 \times 224$ pixels to match the input requirements of the models. Since the selected architectures expect three-channel inputs, the images are converted to RGB format by duplicating the grayscale channel. We employ the Adam optimizer for training the CNN-based models (VGG-16, ResNet-50, and ConvNeXT) with a learning rate of $0.001$. The learning rate is scheduled using the StepLR scheduler with a step size of 5 epochs and a decay factor of 0.1. For transformer-based models (ViT and SWIN, specifically ViT-B-16 and SWIN-B), we use the AdamW optimizer with the same learning rate but apply the CosineAnnealingWarmRestarts scheduler with $T_{\text{max}} = 5$ epochs and $\eta_{\text{min}} = 0$. All models are initialized with pre-trained weights on ImageNet and fine-tuned on our dataset. Training is conducted with a batch size of 32 for 20 epochs. Early stopping is applied based on validation loss to prevent overfitting. All experiments were conducted on a single NVIDIA T4 GPU with 16GB of VRAM.

\vspace{-0.3em}

\subsubsection{Evaluation Metrics}
The performance of the models is assessed using standard classification metrics: accuracy, precision, recall, and F1-score. Accuracy measures the proportion of correctly classified samples. Precision quantifies the ratio of correctly predicted positive samples to total predicted positives, while recall (sensitivity) calculates the ratio of correctly predicted positive samples to total actual positives. The F1-score, the harmonic mean of precision and recall, provides a balanced measure of model performance. While accuracy is computed for the entire dataset, precision, recall, and F1-score are calculated separately for NAT and WAT sources to evaluate class-wise performance.

\vspace{-0.5em}


\subsection{Benchmarking Results}
\subsubsection{Quantitative Results}  
We present the quantitative results of the different models in Table \ref{tab:quantitative} as benchmarking results. The table shows the overall accuracy of the models, as well as the class-wise precision, recall, and F1-score for NAT and WAT sources. ConvNeXT achieves the highest F1-scores for both WAT (0.87) and NAT (0.79), demonstrating its robustness in classifying both morphologies. SWIN-B follows closely, particularly excelling in WAT classification with an F1-score of 0.85, benefiting from its strong recall. Transformer-based models, such as ViT-B-16, also show high recall for WAT sources, suggesting their effectiveness in capturing elongated and complex structures. However, NAT classification remains challenging, with VGG-16 performing the worst (F1-score = 0.60) due to its low recall (0.48). While ResNet-50 provides a balanced performance, it falls short of ConvNeXT and SWIN-B in overall classification effectiveness.

\vspace{-0.6em}

\subsubsection{Qualitative Analysis}
To gain insights into the models' decision-making processes, we visualize the class activation maps (CAMs) using Grad-CAM~\cite{selvaraju2017}. Figure \ref{fig:grad_cam} presents the CAMs generated by all models for a sample NAT and WAT source. The CAMs highlight the regions of the image that contribute most to the model's classification decision. Convolutional neural networks like VGG-16, ResNet-50, and ConvNeXT tend to focus on the core regions of the source, while transformer-based models such as ViT-B-16 and SWIN-B capture the elongated structure of the source more effectively. This observation aligns with the models' performance in classifying WAT sources, where transformers exhibit higher recall compared to CNNs. The CAMs thus provide valuable insights into the models' interpretability, emphasizing the importance of capturing source morphology for accurate classification.

\vspace{-0.6em}

\subsection{Discussion}
The dataset exhibits a notable class imbalance, with 385 WAT and 254 NAT sources, making overall accuracy an inadequate metric for evaluation. Given the small training size (576 samples), F1-score provides a more reliable measure of model effectiveness. Transformer-based models' limitations may stem from the relatively small dataset size, which could hinder their ability to generalize effectively without larger-scale pretraining on similar astronomical images. On the other hand, conventional CNN-based models like VGG-16 and ResNet-50, though outperformed by ConvNeXt, provided baseline insights into the efficacy of simpler architectures for bent AGN detection. These results highlight that CNN-based architectures, particularly ConvNeXT, offer better consistency across both classes, whereas transformers may be advantageous in scenarios requiring high recall, especially for WAT detection.

\vspace{-0.4em}

\section{Licensing and Ethical Considerations}
\label{sec:license}

The dataset is available online at \url{https://doi.org/10.5281/zenodo.15505390} under the Creative Commons Attribution 4.0 International (CC BY 4.0) license. This license permits unrestricted use, distribution, and reproduction in any medium, provided that the original authors and source are credited.

\vspace{-1.8em}
\section{Conclusion}
\label{sec:conclusion}

In this paper, we introduced a novel dataset for the classification of bent radio AGNs, focusing on WAT and NAT sources. With 639 curated radio images, the dataset provides valuable resources for advancing research in astronomical image classification, including detailed data acquisition, processing, and expert validation. We benchmarked various deep learning models, highlighting the strengths of CNN and transformer-based architectures for bent radio AGN classification. This work lays the foundation for future efforts, including expanding the dataset with additional sources from broader radio surveys, contributing to the development of comprehensive resources for astronomical research across multiple domains.

\bibliographystyle{IEEEbib}
\bibliography{strings,refs}
\end{document}